\documentclass[titlepage, prl, amsmath,amssymb,twocolumn, a4]{revtex4-1}

\usepackage{graphicx}
\usepackage{color}
\usepackage{amsmath}
\usepackage{amssymb}

\usepackage{bm}
\usepackage{bbm}

\newcommand{\CO}{\mathbb{C}\otimes\mathbb{O}}

\newcommand{\CLsix}{\mathbb{C}l(6)}
\newcommand{\CLtwo}{\mathbb{C}l(2)}
\newcommand{\CLfour}{\mathbb{C}l(4)}

\newcommand{\oot}{\omega \omega^{\dagger}}
\newcommand{\oto}{\omega ^{\dagger}\omega }

\newcommand{\spacer}{0.2cm}

\begin{document}

\title{Charge quantization from a number operator}

\author{C. Furey}
\affiliation{$ $ \\ Perimeter Institute for Theoretical Physics, Waterloo, Ontario N2L 2Y5, Canada,\\ University of Waterloo, Ontario N2L 3G1, Canada \\ nf252@cam.ac.uk}\pacs{112.10.Dm, 2.60.Rc, 12.38.-t, 02.10.Hh, 12.90.+b}

\begin{abstract}
\noindent We explain how an unexpected algebraic structure, the division algebras, can be seen to underlie a generation of quarks and leptons.  From this new vantage point, electrons and quarks are simply excitations from the neutrino, which formally plays the role of a vacuum state.  Using the ladder operators which exist  within the system, we build a number operator in the usual way.  It turns out that this number operator, divided by 3, mirrors the behaviour of electric charge.  As a result, we see that electric charge is quantized because number operators can only take on integer values.    

Finally, we show that a simple hermitian form, built from these ladder operators, results uniquely in the nine generators of $SU(3)_c$ and $U(1)_{em}$.  This gives a direct route to the two unbroken gauge symmetries of the standard model. 
\end{abstract}

\maketitle


One of the more cherished features of Georgi and Glashow's $SU(5)$ grand unified theory is its ability to succinctly explain the quantization of charge, \citep{su5}, \citep{geng}.  However, the generic features typically bundled into such grand unified theories, e.g. proton decay \citep{sk1}, can often be a heavy burden to carry.  One might then be led to wonder if charge quantization could be found via some \it other \rm mathematical structure, which has fewer strings attached.

Here, we propose one such mathematical structure, whose potential in physics has historically been understated. 
This structure is the set of algebras known as the normed division algebras over the reals.
Strikingly, there exist only four of these algebras: the real numbers, $\mathbb{R}$, the complex numbers, $\mathbb{C}$, the
quaternions, $\mathbb{H}$, and the octonions, $\mathbb{O}$. It can be shown that particle physics relies heavily on the first
three of these algebras.

The real numbers are used almost universally in physics; the complex numbers are central to quantum
theory; the quaternions lead to the Pauli matrices, and are hence tightly entwined with the Lorentz
algebra. In fact, in~\citep{thesis} and \citep{UTI}, it is shown that the complex quaternions can concisely describe all of the Lorentz representations of the standard model: scalars, spinors, four-vectors, and the field strength tensor, in terms of generalized ideals.

But what is to be said for the octonions, $\mathbb{O}$, the fourth, and final division algebra? With $\mathbb{R}$, $\mathbb{C}$,
and $\mathbb{H}$ each undeniably etched into fundamental physics, it is hard not to wonder: is it really the case
that $\mathbb{O}$ has been omitted in nature?

In earlier years, \citep{GGquarks}, \citep{GGstats}, G\"{u}naydin and G\"{u}rsey showed $SU(3)_c$ quark structure in the split octonions.  Furthermore, they showed anti-commuting ladder operators within that model.  Our new results stem from the octonionic chromodynamic quark model of \citep{GGstats}, and are meant to replace the provisional charges of \citep{UTI}.   These findings make a case in support of those who have been long advocating for the existence of a connection between certain non-associative algebras and particle physics, \citep{thesis}-\citep{okubo}.

Using the algebra of the complex octonions, which we will introduce, we expose an intrinsic structure to a generation of quarks and leptons.  This algebraic structure mimics familiar quantum systems, which have a vacuum state acted upon by raising and lowering operators.  In this case, the neutrino poses as the vacuum state, and electrons and quarks pose as the excited states.

With these raising and lowering operators in hand, we are then able to construct a number operator in the usual way,
\begin{equation}N= \sum_i \alpha_i^{\dagger}\alpha_i.
\end{equation}
\noindent It will be seen that $N$ has eigenvalues given by $\{0,1,1,1,2,2,2,3\}$.  At first sight, these eigenvalues might not look familiar, that is, until they are divided by 3.  $N/3$ has eigenvalues $\{0,\hspace{.5mm}\frac{1}{3},\hspace{.5mm}\frac{1}{3},\hspace{.5mm}\frac{1}{3}, \hspace{.5mm}\frac{2}{3},\hspace{.5mm}\frac{2}{3},\hspace{.5mm}\frac{2}{3},\hspace{.5mm}1\}$, which can now be recognized as the electric charges of a neutrino (or anti-neutrino), a triplet of anti-down-type quarks, a triplet of up-type quarks, and a positron.   We will then define our electric charge, $Q$, as
\begin{equation}\label{Q}
Q\equiv\frac{N}{3}.
\end{equation}
\noindent  As $N$ must take on integer values, $Q$ must be quantized.  

As we will show, the remaining states within a generation are related to these particles by complex conjugation, and hence are acted upon by $-Q^*$ in the usual way. 

Ours is certainly not the first instance where G\"{u}naydin and G\"{u}rsey's model has been adapted.  As an extension of their model, \citep{dixon}, \citep{matter}, Dixon describes electric charge  as a mix of quaternionic and octonionic objects.    It would be interesting to see if a ladder system could be found, which alternately gives Dixon's $Q$ as a number operator.   Readers are encouraged to see \citep{dixon}, \citep{matter}, or other examples of his extensive work.

Since the time of first writing, more octonionic chromo-electrodynamic models have been found.  Most noteworthy of all were three papers written in the late 1970s, \citep{Grass}, \citep{wow} and \citep{wow2}, which could also be considered as extensions of G\"{u}naydin and G\"{u}rsey's model,~\citep{GGstats}.  In these papers, the authors use two separate ladder systems:  system (a) fits with the octonionic ladder operators of \citep{GGstats}, and system (b) is introduced as quaternionic.  By combining the two systems, they describe the electric charge generator not as a number operator, but as the difference  between the number operators of the two systems.  References \citep{Grass}, \citep{wow}, and \citep{wow2} are important papers, worth careful reconsideration by the community. 

Our results differ from earlier versions in that we will be constructing a generation of quarks and leptons explicitly as \it minimal left ideals \rm of a Clifford algebra, generated by the complex octonions.  In doing so, we will use just a single octonionic ladder system, with its complex conjugate.  This in turn allows us to define electric charge more simply as $Q=N/3$, thereby exposing a more direct route to the two unbroken gauge symmetries of the standard model.  Furthermore,  our formalism naturally relates particles and anti-particles using only the complex conjugate, $i\mapsto -i$, which is not a feature of these earlier models.  Finally, as our generation of quarks and leptons will be constructed from Clifford algebra elements, not column vectors, we will then be free to model weak isospin, using \it right multiplication \rm of this same Clifford algebra onto these minimal left ideals.



\noindent \bf Acquaintance with  $\mathbb{C}\otimes\mathbb{O}$. \rm  The complex octonions are not a tool commonly used in physics, so we introduce them here.

\begin{figure}[h!]
\includegraphics[width=6cm]{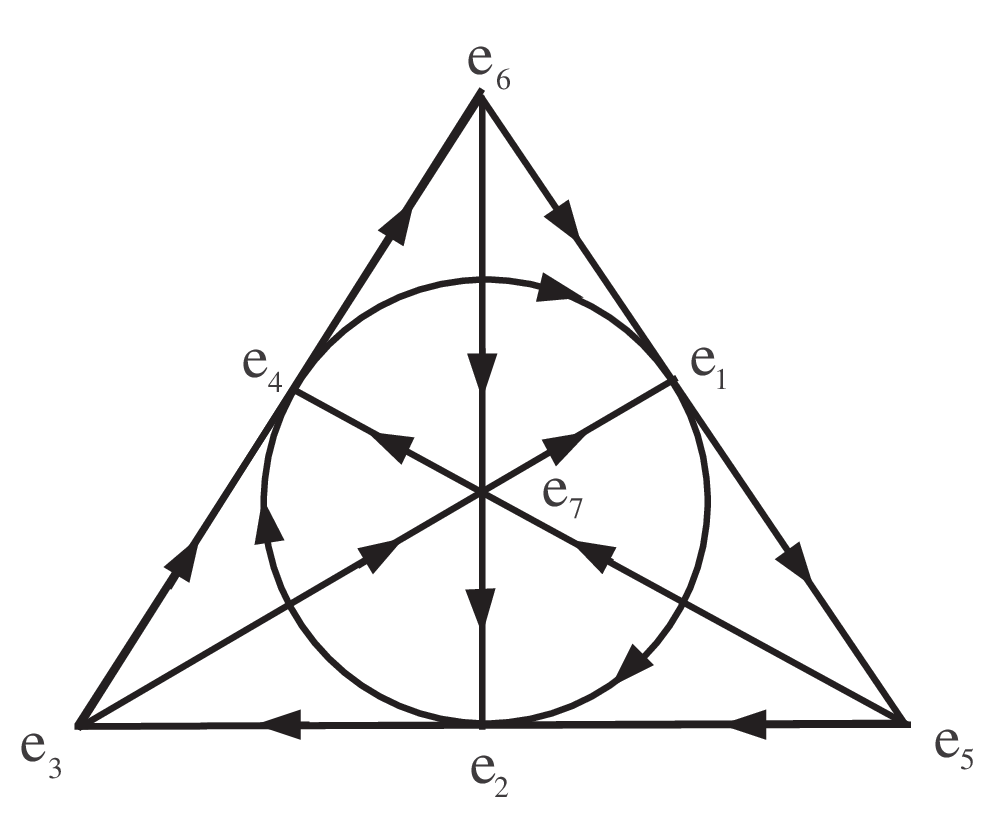}
\caption{\label{fano}
Multiplication of octonionic imaginary units}
\end{figure}

Any element of $\mathbb{C} \otimes \mathbb{O}$ can be written as $ \sum_{n=0}^7 A_n e_n $, where the $A_n$ are complex coefficients.  The $e_n$ are octonionic imaginary units $\left(e_n^2=-1\right)$, apart from $e_0=1$, which multiply as per Figure~\ref{fano}.  The complex imaginary unit, $i,$ commutes with each of the octonionic $e_n$.

Any three imaginary units on a directed line segment in Figure~\ref{fano} act as if they were a triplet of Pauli matrices, $\sigma_m$.  (More precisely, they behave as $-i\sigma_m$.)  For example, $e_6e_1 =-e_1e_6= e_5,$ $e_1e_5=-e_5e_1=e_6,$ $e_5e_6=-e_6e_5=e_1,$ $e_4e_1=-e_1e_4=e_2$, etc.  It is indeed true that the octonions form a non-associative algebra, meaning that the relation $(ab)c =a(bc)$ does not always hold.   The reader can check this by finding three imaginary units, which are not all on the same line segment, and substituting them as in $a$, $b$, and $c$.  For a more detailed introduction of $\mathbb{O}$ see~\citep{baez}, \citep{conway}, \citep{okubo}.

Finally, we define three notions of conjugation on an element $a$ in $\mathbb{C}\otimes\mathbb{O}$.  The \it complex conjugate \rm of $a$, denoted $a^*$, maps the complex $i\mapsto -i$, as would be expected.  The \it octonionic conjugate \rm of $a$, denoted $\tilde{a}$, takes each of the octonionic imaginary units $e_n\mapsto -e_n$ for $n=1,\dots 7$.  That which we will call the \it hermitian conjugate \rm of $a$, denoted $a^{\dagger}$, performs both of these maps simultaneously, $i \mapsto -i$ and $e_n \mapsto -e_n$ for $n=1,\dots 7$.  The octonionic conjugate and the hermitian conjugate each reverse the order of multiplication, as is familiar from the hermitian conjugate of a product of matrices.

\noindent \bf A system of ladder operators. \rm  Upon some exploration, one finds a system of ladder operators within the complex octonions.  Consider $\alpha_1 \equiv \frac{1}{2}\left(-e_5+ie_4\right)$,  $\alpha_2 \equiv \frac{1}{2}\left(-e_3+ie_1\right),$ and $\alpha_3 \equiv \frac{1}{2}\left(-e_6+ie_2\right)$, similar to that defined in~\citep{GGquarks}.  For all $f$ in $\mathbb{C}\otimes\mathbb{O}$, and assuming right-to-left multiplication, these three lowering operators obey the anticommutation relations
\begin{equation}\label{alpha}
\{\alpha_i, \alpha_j\}f =\alpha_i(\alpha_jf)+\alpha_j (\alpha_if) = 0 
\end{equation}
\noindent for all $i,j = 1,2,3.$  The above can be seen as a generalization of the system in~\citep{GGstats}.  That is, \citep{GGstats} is recovered by restricting the general $f$ in $\mathbb{C}\otimes\mathbb{O}$ to $f=1$.  

In another slight deviation from \citep{GGstats}, we define raising operators as $\alpha_1^{\dagger} = \frac{1}{2}\left(e_5+ie_4\right)$,  $\alpha_2^{\dagger} = \frac{1}{2}\left(e_3+ie_1\right),$ and $\alpha_3^{\dagger} = \frac{1}{2}\left(e_6+ie_2\right)$, which obey
\begin{equation}
\{\alpha_i^{\dagger}, \alpha_j^{\dagger}\}f = 0 \hspace{0.7cm}\textup{for all} \hspace{0.2cm} i,j = 1,2,3.
\end{equation}

\noindent We finally also have 
\begin{equation}
\{\alpha_i,\alpha_j^{\dagger}\}f=\delta_{ij}f \hspace{0.7cm} \textup{for all} \hspace{0.2cm}i,j = 1,2,3.
\end{equation}

With the purpose simplifying notation, we will now implicitly assume all multiplication to be carried out starting at the right, and moving to the left, as was shown in equation~(\ref{alpha}).  That is, we will now not write these brackets in explicitly.   Furthermore,  we will now be concerned only with operators, such as the $\alpha_i$, as opposed to the object $f$.  This being the case, it will now be understood that all equations will hold over all $f$ in $\CO$, even though $f$ will not be mentioned explicitly.    For example, we will now write equation~(\ref{alpha}) simply as
\begin{equation}
\{\alpha_i, \alpha_j\} =\alpha_i\alpha_j+\alpha_j \alpha_i = 0 \hspace{1cm}\textup{for all}\hspace{0.2cm} i,j = 1,2,3.
\end{equation}

Incidentally, these operators acting on $f$ may be viewed as $8\times8$ complex matrices acting on $f$, an eight-complex-dimensional column vector.  Taking into account the above paragraph, our equations from here on in can be considered as relations only between the matrices.

\noindent \bf Complex conjugation's analogue. \rm   Under complex conjugation, we find an analogous ladder system.  Consider $\alpha_1^* = \frac{1}{2}\left(-e_5-ie_4\right)$,  $\alpha_2^* = \frac{1}{2}\left(-e_3-ie_1\right),$ and $\alpha_3^* = \frac{1}{2}\left(-e_6-ie_2\right)$.  These three lowering operators obey the anticommutation relations
\begin{equation}
\{\alpha_i^*, \alpha_j^*\} = 0 \hspace{0.7cm}\textup{for all}\hspace{0.2cm} i,j = 1,2,3.
\end{equation}

We define raising operators as $\tilde{\alpha}_1 = \frac{1}{2}\left(e_5-ie_4\right)$,  $\tilde{\alpha}_2 = \frac{1}{2}\left(e_3-ie_1\right),$ and $\tilde{\alpha}_3 = \frac{1}{2}\left(e_6-ie_2\right)$, which obey
\begin{equation}
\{\tilde{\alpha}_i, \tilde{\alpha}_j\} = 0 \hspace{0.7cm}\textup{for all} \hspace{0.2cm} i,j = 1,2,3.
\end{equation}

\noindent Finally, we have also
\begin{equation}\label{alphastar}
\{\alpha_i^*,\tilde{\alpha}_j\}=\delta_{ij}\hspace{0.7cm} \textup{for all} \hspace{0.2cm}i,j = 1,2,3.
\end{equation}

Using these ladder operators, we will now build \it minimal left ideals, \rm which can be seen to mimic the set of quarks and leptons of the standard model.  

\noindent \bf Minimal left ideals. \rm  Intuitively speaking, an \it ideal \rm is a special subspace of an algebra because it is robust under multiplication.  For this reason, ideals are well suited to describe particles persisting under evolution and transformation.

Given an algebra, $A$, a \it left ideal, \rm $B$, is a subalgebra of $A$ whereby $a b$ is in $B$ for all $b$ in $B$, and for any $a$ in $A$.  That is, no matter which $a$ we multiply onto $b$, the new product, $b'\equiv ab$, cannot leave the subspace $B$.  It is easy to see how $b'\equiv ab$ could easily describe, for example, a particle $b$ undergoing a transformation $a$.

A  \it minimal left ideal \rm is a left ideal which contains no left ideals other than $\{ 0 \} $ and itself.  In other words, it has no non-trivial ideals inside it.

In this article, we are proposing to represent quarks and leptons using minimal left ideals within our space of octonionic operators:  that is, within the space of the $\alpha_i$, $\alpha^{\dagger}_j$, and their products.  A pair of these ideals, $S^u$ and $S^d$, will be introduced below.  Readers wishing to confirm the construction may consult~\citep{Gen} for an explanation of how left multiplication of $\CO$ on itself gives a representation of the 64-complex-dimensional Clifford algebra $\CLsix$.    The review, \citep{ablam}, then lucidly describes the construction of minimal left ideals in Clifford algebras via \it Witt decomposition.  \rm   (For an alternate phase space perspective on the real Clifford algebra $Cl(6)$, see~\citep{zen}.)

From our first ladder system, we define
\begin{equation}\label{omega}\begin{array}{c}
\omega  \hspace{1mm}\equiv \hspace{1mm} \alpha_1\alpha_2\alpha_3, \vspace{\spacer} \\
\omega^{\dagger}  \hspace{1mm}\equiv \hspace{1mm}\alpha_3^{\dagger}\alpha_2^{\dagger}\alpha_1^{\dagger},
\end{array}\end{equation}
\noindent which lead to the identies $\omega ^{\dagger}\oot = \omega ^{\dagger}$ and $\oot\oot = \oot$.

The eight-complex-dimensional minimal left ideal for the first ladder system is given by
\begin{equation}\begin{array}{lrlcll}  \label{su}
 &S^{u}  \equiv&&&& \vspace{\spacer}\\
&& &\mathcal{V}  \hspace{1mm} \omega \omega ^{\dagger}   & & \vspace{\spacer}\\

   & +  \hspace{1mm}\bar{\mathcal{D}}^{\textup{r}} \hspace{1mm}  \alpha_{1}^{\dagger}\oot    & + &   \hspace{1mm}\bar{\mathcal{D}}^{\textup{g}}  \hspace{1mm} \alpha_{2}^{\dagger}\oot    & + & \hspace{1mm}\bar{\mathcal{D}}^{\textup{b}}  \hspace{1mm} \alpha_{3}^{\dagger}\oot   \vspace{\spacer}\\

    & +\hspace{1mm}\mathcal{U}^{\textup{r}} \hspace{1mm}  \alpha_{3}^{\dagger}\alpha_{2}^{\dagger} \oot   & + & 
\hspace{1mm}\mathcal{U}^{\textup{g}} \hspace{1mm}  \alpha_{1}^{\dagger}\alpha_{3}^{\dagger} \oot  & + &  \hspace{1mm}\mathcal{U}^{\textup{b}} \hspace{1mm}  \alpha_{2}^{\dagger}\alpha_{1}^{\dagger} \oot  \vspace{\spacer} \\

  & & +&\hspace{1mm}\mathcal{E}^{+}  \hspace{.7mm}  \alpha_{3}^{\dagger}\alpha_{2}^{\dagger}\alpha_{1}^{\dagger} \oot,  & &
\end{array}\end{equation}
\noindent where $\mathcal{V},$  $\bar{\mathcal{D}}^{\textup{r}}, \dots \mathcal{E}^{+} $ are 8 suggestively named complex coefficients.  

As 
\begin{equation} \alpha_i\hspace{1mm}\oot = 0 \hspace{1cm} \forall i,
\end{equation}
\noindent $\oot$ plays the role of the vacuum state, where the term \it vacuum \rm is used loosely.  Readers may recognize the similarity between $S^u$ and a Fock space.

The conjugate system analogously leads to

\begin{equation}\begin{array}{lrlcll}  \label{sd}
&S^{d} \equiv&&&& \vspace{\spacer}\\

 & & &\bar{\mathcal{V}}  \hspace{1mm} \oto   & & \vspace{\spacer}\\

   &   \hspace{1mm}\mathcal{D}^{\textup{r}} \hspace{1mm}  (-\alpha_{1}\oto)    & + &   \hspace{1mm}\mathcal{D}^{\textup{g}}  \hspace{1mm} (-\alpha_{2}\oto)    & + & \hspace{1mm}\mathcal{D}^{\textup{b}}  \hspace{1mm}(-\alpha_{3}\oto)   \vspace{\spacer}\\

    & +\hspace{1mm}\bar{\mathcal{U}}^{\textup{r}} \hspace{1mm}  \alpha_{3}\alpha_{2} \oto  & + & 
\hspace{1mm}\bar{\mathcal{U}}^{\textup{g}} \hspace{1mm}  \alpha_{1}\alpha_{3} \oto  & + &  \hspace{1mm}\bar{\mathcal{U}}^{\textup{b}} \hspace{1mm}  \alpha_{2}\alpha_{1} \oto \vspace{\spacer} \\

  & & +&\hspace{1mm}\mathcal{E}^{-}  \hspace{.7mm} \alpha_1\alpha_{2}\alpha_{3} \oto,  & &
\end{array}\end{equation}
\noindent where $\bar{\mathcal{V}},$  $\mathcal{D}^{\textup{r}}, \dots \mathcal{E}^{-} $ are eight complex coefficients.   

This new ideal, (\ref{sd}), is linearly independent from the first, (\ref{su}), in the space of octonionic operators.   Clearly, the two are related via the complex conjugate, $i \mapsto -i$.  In fact, the complex conjugate is \it all \rm that is needed in order to map particles into anti-particles, and vice versa.  This was a feature in the models of~\citep{GGstats}, \citep{Gen}, and also in models of left- and right-handed Weyl spinors,~\citep{thesis}, \citep{UTI}.  It was not a feature in \citep{Grass}, \citep{wow}, \citep{wow2}, \citep{dixon}, or \citep{matter},  where an additional quaternionic algebra was implemented in order to obtain $S^d$.

The Clifford algebra $\CLsix$ is known to have just a single 8-complex-dimensional irreducible representation, as in $S^u$, above.  In this paper, we will none-the-less be including the conjugate ideal, $S^d$, in anticipation of future work, which will combine $S^u$ and $S^d$ into a single irreducible representation under $\CLsix \otimes \CLtwo$.  (Later on, we will then consider $\CLsix \otimes \CLfour$, suggesting a connection to the Pati-Salam model.)  Unlike in the earlier literature, this additional factor of $\CLtwo$ will originate from right multiplication of our octonionic operators on these ideals, as mentioned at the end of this text.  This $S^u+S^d$ form is motivated by the \it complex multiplicative action, \rm described in~\citep{thesis}.

As a final note, we point out that another interesting way to obtain anti-particles could be to use the conjugate $\dagger$, instead of $*$.  In that case, the two minimal left ideals would not be entirely linearly independent from each other.  That is, we would find a special Majorana-like property unique to the neutrino:  $\left(\oot\right)^{\dagger} = \oot$.

\noindent \bf Ladders to the unbroken symmetries. \rm  Having obtained these minimal left ideals, we would now like to know how they transform, so as to justify the labels we gave to their coefficients in equations (\ref{su}) and (\ref{sd}).  It so happens that a very simple form leads uniquely  to the generators of the two unbroken gauge symmetries of the standard model,  $SU(3)_c$ and $U(1)_{em}$.  We will find these generators, and then apply them to our minimal left ideals.

Consider $\alpha \equiv c_1\alpha_1 + c_2\alpha_2 +c_3\alpha_3$ and $\alpha' \equiv c'_1\alpha_1 + c'_2\alpha_2 +c'_3\alpha_3$, where the $c_i$ and $c'_j$ are complex coefficients.  We can then build hermitian operators, $\mathcal{H}$, of the form 
\begin{equation}
\mathcal{H}\equiv \alpha'^{\dagger}\alpha  +   \alpha^{\dagger}\alpha'.
\end{equation}
\noindent Taking the most general sum of these objects results in nine hermitian operators:
\begin{equation}\label{unique}
\sum_{\mathcal{H}} \mathcal{H}\hspace{1.2mm} = \hspace{1.2mm}r_0\hspace{0.7mm}Q+\hspace{0.7mm}\sum_{i=1}^8 r_i\Lambda_i,
\end{equation}
\noindent where $r_0$ and $r_i$ are real coefficients.  $Q$ is our electromagnetic generator from equation~(\ref{Q}), and the eight $\Lambda_i$ can be seen to generate $SU(3)_c$.  Indeed, these $\Lambda_i$ coincide with those introduced in~\citep{GGstats} (which generate a subgroup of the octonionic automorphism group, $G_2$).  Readers are encouraged to also see~\citep{thesis}, where these nine generators are further identified as symmetry generators preserving ladder operator structure, otherwise known as \it unitary MTIS symmetries. \rm

The result of equation~(\ref{unique}) is worth emphasizing.  That is, the simple form, $\sum_{\mathcal{H}} \mathcal{H}$ leads \it uniquely \rm to the generators of the two unbroken gauge symmetries of the standard model.

In terms of ladder operators, the $SU(3)_c$ generators take the form
\begin{equation}\begin{array}{ll}\label{liealg}
\Lambda_1= -\alpha_2^{\dagger}\alpha_1    -\alpha_1^{\dagger}\alpha_2           &                \Lambda_2=i\alpha_2^{\dagger}\alpha_1 -i\alpha_1^{\dagger}\alpha_2       \vspace{\spacer}\\
\Lambda_3=\alpha_2^{\dagger}\alpha_2  -\alpha_1^{\dagger}\alpha_1
& \Lambda_4=-\alpha_1^{\dagger}\alpha_3-\alpha_3^{\dagger}\alpha_1        \vspace{\spacer}
\\ \Lambda_5=-i\alpha_1^{\dagger}\alpha_3 +i\alpha_3^{\dagger}\alpha_1& \Lambda_6=-\alpha_3^{\dagger}\alpha_2-\alpha_2^{\dagger}\alpha_3
\vspace{\spacer}
\\ \Lambda_7=i\alpha_3^{\dagger}\alpha_2-i\alpha_2^{\dagger}\alpha_3 & \Lambda_8=  -\frac{1}{\sqrt{3}}\left[\alpha_1^{\dagger}\alpha_1  + \alpha_2^{\dagger}\alpha_2  -2\alpha_3^{\dagger}\alpha_3\right],
\end{array}\end{equation}

\noindent all eight of which can be seen to commute with $Q$, and its conjugate.

Now, we take the minimal left ideal, $S^u$, to transform as 
\begin{equation}\label{transfspin}
e^{i\sum\mathcal{H}} \hspace{1mm}S^u \hspace{1mm}e^{-i\sum\mathcal{H}} =e^{i\sum\mathcal{H}} \hspace{1mm}S^u  ,
\end{equation}
\noindent  where the equality holds because $\omega^{\dagger}\alpha_i^{\dagger} = 0$ for all $i$.   

We now identify the subspaces of $S^{u}$ by specifying their electric charges with respect to $U(1)_{em}$, and also which irreducible representation they belong to under $SU(3)_c$.  Clearly, $i$, $j$ and $k$ are meant to be distinct from each other in any given row.

\begin{equation}\begin{array}{cccc}\label{upcharge}
\hspace{.75cm}\underline{Q}\hspace{.75cm}   &    \hspace{.75cm}\boldmath{\underline{\Lambda}} \hspace{.75cm}&\hspace{.75cm} \underline{S^u} \hspace{.75cm}& \hspace{.75cm}\underline{\textup{ID}} \hspace{.75cm}\vspace{6mm}\\

0 & 1 & \oot & \nu \hspace{1mm} \left(\textup{or}\hspace{1mm} \bar{\nu}\right)   \vspace{\spacer}\\

1/3 & \bar{3} & \alpha_i^{\dagger}\oot & \bar{d}_i  \vspace{\spacer}\\

2/3 & 3 &  \alpha_i^{\dagger}\alpha_j ^{\dagger}\oot & u_k \vspace{\spacer}\\

1 & 1 & \alpha_i^{\dagger}\alpha_j ^{\dagger}\alpha_k^{\dagger}\oot & e^+    \vspace{\spacer}
\end{array}\end{equation}

\noindent So, here we identify a neutrino, $\nu$, (or antineutrino, $\bar{\nu}$), three anti-down type quarks, $\bar{d}_i$, three up-type quarks, $u_k$, and a positron, $e^+$.

As the minimal left ideal, $S^d$, is related to $S^u$ by complex conjugation, we then see that it transforms as 
\begin{equation}\label{transfspin}
e^{-i\sum\mathcal{H}^*} \hspace{1mm}S^d \hspace{1mm}e^{i\sum\mathcal{H}^*} =e^{-i\sum\mathcal{H}^*} \hspace{1mm}S^d  ,
\end{equation}
\noindent where the equality holds because $\omega \alpha_i =0$ for all $i$.  This leads to the table below.  

\begin{equation}\begin{array}{cccc}\label{downcharge}
\hspace{.75cm}-\underline{Q}^*\hspace{.75cm}   &    \hspace{.75cm}\boldmath{-\underline{\Lambda}^*} \hspace{.75cm}&\hspace{.75cm} \underline{S^{d}} \hspace{.75cm}& \hspace{.75cm}\underline{\textup{ID}} \hspace{.75cm}\vspace{6mm}\\

0 & 1 & \oto&\bar{\nu} \hspace{1mm} \left(\textup{or}\hspace{1mm} \nu\right)   \vspace{\spacer}\\

-1/3 & 3 & \alpha_i\oto & d_i\vspace{\spacer}\\

-2/3 & \bar{3} &  \alpha_i\alpha_j \oto & \bar{u}_k \vspace{\spacer}\\

-1 & 1 & \alpha_i\alpha_j \alpha_k\oto   & e^-   \vspace{\spacer}
\end{array}\end{equation}

\noindent Here, we identify an antineutrino, $\bar{\nu}$, (or a neutrino, $\nu$), three down-type quarks, $d_i$,  three anti-up type quarks, $\bar{u}_k$, and the electron, $e^-$.

\begin{figure}[h!]
\includegraphics[width=9.1cm]{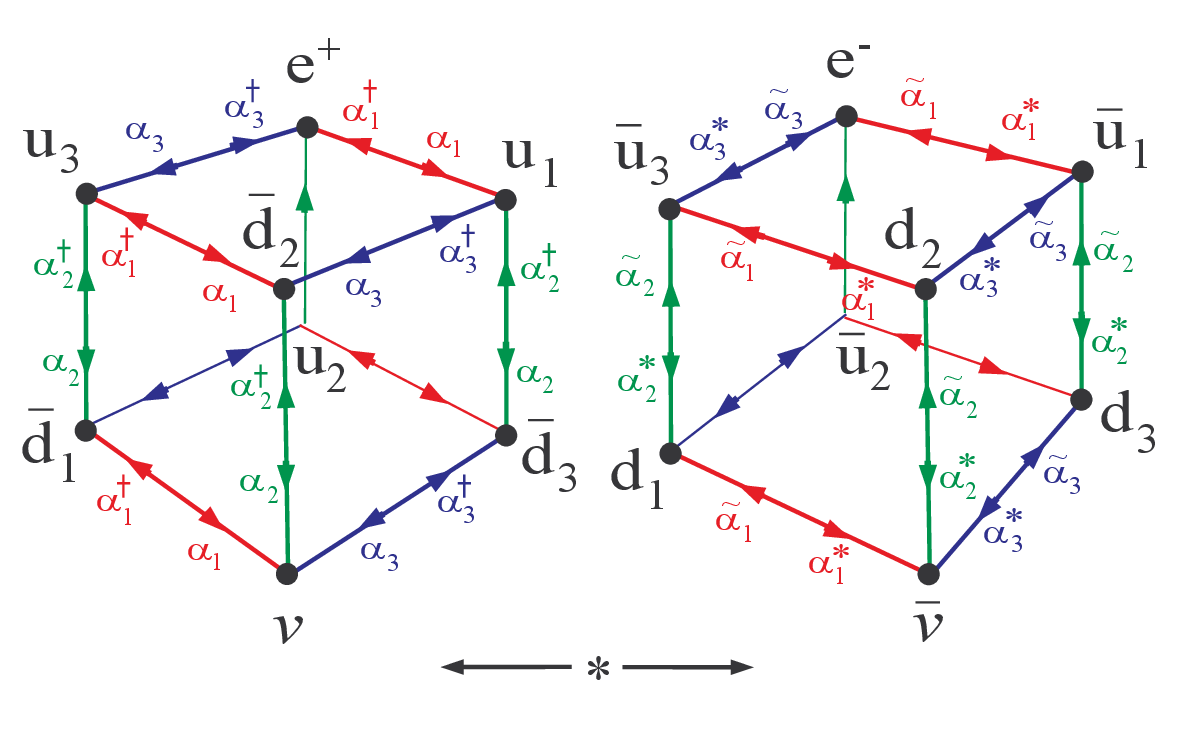}
\caption{\label{cube}  A full generation represented by cubes $S^u$ (left) and $S^d$ (right).  Quark and electron states may be viewed as excitations from the neutrino or anti-neutrino.  As the ``vacuum" represents the neutrino, and not the zero particle state, this model does not constitute a composite model in the usual sense. }
\end{figure}

We have now shown a pair of conjugate ideals, which behave under $su(3)_c$ and $u(1)_{em}$ as does a full generation of the standard model.   These are summarized in Figure (\ref{cube}).

\noindent \bf A signal from W bosons. \rm  Perhaps unexpectedly, it turns out that $S^u$  packages all of the isospin up-type states together,  and $S^d$ packages all of the down-type states together.  This is of course, if one goes ahead and makes an assumption about the placement of $\nu$ into $S^u$ and $\bar{\nu}$ into $S^d$.

We point out that $\omega$ is negatively charged, and converts isospin up particles into isospin down, via \it right \rm multiplication on $S^u$.   It thereby exhibits features of the $W^-$ boson.  Similarly, $\omega^{\dagger}$ is positively charged, and converts isospin down particles into isospin up, via \it right \rm multiplication on $S^d$.  In doing so, it exhibits features of the $W^+$ boson.  

Other characteristics of the W bosons do not appear at the level of this article.   For example, there is nothing to specify that these candidate bosons act only on left-handed particles.  Some steps in this direction have been made in Chapter 7 of~\citep{thesis}.


\noindent \bf Conclusion. \rm  Using only the complex octonions acting on themselves, we were able to recover a number of aspects of the standard model's structure.

First of all, we found that a simple hermitian form led \it uniquely \rm to the two unbroken gauge symmetries of the standard model, $su(3)_c$ and $u(1)_{em}$.  This new $U(1)_{em}$ generator, $Q$, happens to be proportional to a number operator, thereby suggesting an unexpected resolution to the question:  Why is electric charge quantized?

Then, using octonionic ladder operators, we have built a pair of minimal left ideals, which is found to transform under these unbroken symmetries  as does a generation of quarks and leptons.

\medskip

 If the algebra of the complex octonions
is \it not \rm behind the structure of the standard model, it is
then a striking coincidence that $su(3)_c$ and $u(1)_{em}$ both
follow readily from its ladder operators. 

\smallskip

\noindent \bf Acknowledgements. \rm Thank you to G. Fiore,  B. Foster,  M. G\"{u}naydin, C. Tamarit, and especially L. Boyle and A. Kempf  for their feedback and support.

Quite some time ago, an interim PhD supervisor, L. Smolin, suggested looking at anti-down quarks instead of down quarks in the context of~\citep{UTI}.  This advice proved useful in that it made the electric charge generator immediately recognizable within the analysis.

This research was supported by the Templeton Foundation, and also in part by Perimeter Institute for Theoretical Physics. Research at Perimeter Institute is supported by the Government of Canada through Industry Canada and by the Province of Ontario through the Ministry of Research and Innovation.

\end{document}